# Anisotropic magnetodielectric coupling behavior of Ca$_3$Co$_{1.4}$Rh$_{0.6}$O$_6$ due to geometrically frustrated magnetism

Tathamay Basu, Kartik K Iyer, Kiran Singh,$^@$ K. Mukherjee, P.L. Paulose and E.V. Sampathkumaran

*Tata Institute of Fundamental Research, Homi Bhabha Road, Colaba, Mumbai 400005, India*

**Abstract**

We have investigated the magnetic, dielectric and magnetodielectric (MDE) behavior of a geometrically frustrated spin-chain system, Ca$_3$Co$_{1.4}$Rh$_{0.6}$O$_6$, in the single crystalline form for different orientations. The results bring out that the magnetic behavior of this compound is by itself interesting in the sense that this compound exhibits an anisotropic glassy-like magnetic behavior with a huge frequency ($\nu$) dependence of ac susceptibility ($\chi$) peak for an orientation along the spin-chain in the range 30-60 K; this behavior is robust to applications of large external magnetic fields ($H$) unlike in canonical spin-glasses. The temperature dependence of dielectric constant also shows strong $\nu$-dependence with similar robustness to $H$. The isothermal $H$-dependent dielectric results at low temperatures establishes anisotropic MDE coupling. It is intriguing to note that there is a 'step' roughly at one-third of saturation values as in the case of isothermal magnetization curves for same temperatures (for orientation along spin-chain), a correlation hitherto unrealized for geometrically frustrated systems.

It is well-known that the geometrical frustration plays an important role in inducing magnetoelectric coupling, a topic of current interest due to possible applications [1, 2]. In this letter, we bring out magnetodielectric (MDE) coupling induced by geometrically frustrated anisotropic glassy-like magnetism in a Ca-Co-Rh-O system, crystallizing in $K_4CdCl_6$-type rhombohedral structure [see, for instance, Refs. 3-5]. It is well-known that the compound $Ca_3CoRhO_6$ contains ferromagnetic chains [6] of $Co^{2+}$ (high-spin at trigonal prismatic site) and $Rh^{4+}$ (low spin at octahedral site) ions running along $c$-axis, separated by Ca ions. Various studies on polycrystals of $Ca_3CoRhO_6$ confirmed that there are two magnetic transitions, one near ($T_N= T_1=$) 90 K and the other near ($T_2=$) 30 K. As a consequence of geometrical frustration, between $T_1$ and $T_2$, one-third of the chains (that is, the ones present at the centre of the hexagon) are believed to be incoherent, while other chains at the corners of the hexagon are coupled antiparallel. This structure is called "partially disordered antiferromagnetism (PDA)" [6]. Below $T_2$, a complex spin-glass-like phase appears. Readers may see the references 6-23 for various other properties. Barring an initial preliminary report [7], there has been no attempt to prepare single-crystalline $Ca_3CoRhO_6$. We report here the results of ac and dc magnetization ($M$) as well as of dielectric permittivity studies on single-crystalline $Ca_3CoRhO_6$.

Needle-shaped hexagonal single crystals (see supplementary material [24]) were grown from a molten $K_2CO_3$ flux and the composition turned out to be $Ca_3Co_{1.4}Rh_{0.6}O_6$ as in Ref. 7. For experimental details of dc $M$, ac susceptibility ($\chi$) and dielectric measurements, see our Refs. 4 and 10.

The results of temperature ($T$) dependence of dc $M$ measured with two different fields ($H=$ 100 Oe and 5 kOe) for two orientations, $H//c$ and $H\perp c$, are shown in figure 1a-b. [$c$-axis is the long rod direction along which the crystal grows, as evidenced from $M(H)$ data also, presented below]. For $H//c$, the behavior of dc $\chi$, both for zero-field-cooled (ZFC) and field-cooled (FC) conditions, is qualitatively similar to that known for polycrystals (for both the fields) of $Ca_3CoRhO_6$ [10], except a difference in the values $T_1$ and $T_2$; that is, in the $\chi(T)$ curve, following a gradual increase with decreasing $T$, the upturn due to $T_1$ occurs at 60 K, instead of 90 K; a sharp drop at $T_2$ occurs around 25 K instead of taking place near 35 K. This reduction in $T_1$ and $T_2$ in single crystals is attributed to the deviation in the stoichiometry [7]. For the orientation $H\perp c$, the magnitude of $\chi$ is more than an order of magnitude smaller, which establishes strongly anisotropic nature of magnetism; this could be consistent with strong intrachain ferromagnetic interaction [8] as well as with fact that the rod direction is the $c$-axis, which is the easy axis of magnetization. The features appearing near $T_1$ and $T_2$ are so weak that it may not be intrinsic to this orientation, but due to a small misalignment of the crystal towards the orientation $H//c$. In support of this, there is no signal for this orientation in ac $\chi$ (see below). In short, this compound exhibits anisotropic magnetic ordering.

In figures 1c and 1d, we show isothermal $M$ behavior below $T_1$. For $H//c$, for instance, for 30 and 40 K, there is a distinct evidence for a step in the range 20-40 kOe at one-third of saturation magnetization, characteristic of PDA structure of a triangular lattice, as seen in polycrystals [9]. For 30 K, a small hysteresis loop around this critical field is observed (see inset of figure 1c), which implies disorder-broadened first-order nature of the metamagnetic transition. However, this step is absent below $T_2$ (see the plot for 2K in figure 1c) as in polycrystals; a distinct hysteretic loop is observed, which is not inconsistent with the spin-glass-like nature of magnetism below $T_2$. Above $T_1$, $M(H)$ curve tends towards paramagnetic behavior (see figure 1d for 62K). The $M(H)$ step behavior is totally different for $H\perp c$; $M$ is found to vary linearly with $H$ for all temperatures without hysteresis (see figure 1e); the magnetic-moment values for a given field are significantly smaller with respect to those for $H//c$ consistent with anisotropic magnetism. Incidentally, all these features are in broad agreement with those seen for oriented polycrystals [9], thereby supporting evidence for the fact that the rod axis is the $c$-axis.

The real ($\chi'$) and imaginary ($\chi''$) parts of ac $\chi$ obtained with various frequencies (ν) and different magnetic fields for $H//c$ are shown in figure 2. Both these components show intense peaks in the vicinity of magnetic transitions. The magnitude of the shift of the peak temperature ($T_p$) is huge with ν, that is, about 10 K for a variation of ν from 1.3 Hz to 1.3 kHz. Appearance of similar features in $\chi''$ favors spin-glass-like dynamics. Since similar shift of $T_p$ with ν in ac χ was observed even in stoichiometric polycrystals [10], we presume that the observed spin-glass-like anomaly is not due to disorder arising from off-stoichiometry of the single crystal. The plot of lnν versus inverse of $T_p$ (peak temperature in $\chi''$) is linear (inset of figure 2), suggesting that Arrhenius relation (ν= $\nu_0 \exp(E_a/k_B T_p)$, $\nu_0$= pre-exponential factor, $E_a$= activation energy) is applicable. The value of $E_a$ turns out to be about 990 K. In order to look for other characteristics of spin-glasses, we performed 'memory' experiments [25]. For this purpose, we obtained ZFC (warming) dc χ curves after waiting at various temperatures for some hours below 60 K while cooling, but we could not find any 'local dip' at the waiting temperatures with respect to those obtained without waiting. One can offer two explanations for the absence of the dip: (i) The spin-glass behavior is restricted to one-third of the total chains and the expected 'dip' being is being masked by the dominant behavior of other magnetically-ordered spin-chains; and (ii) the spin-dynamics arising from geometrical frustration is so complex and slow that the dip may be smeared. It is worth noting that we could not observe any feature in the third-order harmonic (expected for spin-glasses) in ac χ at the magnetic transitions. This observation suggests that this compound may not be classified along with conventional spin-glasses. For such reasons, we describe the magnetic behavior as 'spin-glassy-like', rather than calling it 'canonical spin-glass'. Turning to the behavior for the orientation $H \perp c$, both $\chi'$ and $\chi''$ are featureless in the entire $T$-range, without any evidence for a peak at the magnetic transitions. This finding reveals strongly anisotropic nature of glassy-like magnetic behaviour. This finding is of significance considering great interest in 'anisotropic spin-glass freezing' [26]. See, supplementary material [24] for additional support from isothermal remnant magnetization, $M_{IRM}$. Finally, with respect to the ac χ behaviour in the presence of dc magnetic fields of 10 and 50 kOe (shown in figures 2b and 2c respectively), surprisingly, the strongly ν-dependent nature observed (for $H//c$) is remarkably invariant under magnetic fields, though the data become noisy with increasing fields. In other words, $T_p$ for a particular frequency remains the same for various values of $H$. (Two vertical dotted lines are drawn in figure 2 to bring out this point). This behaviour makes the glassy behavior of this compound a peculiar one.

Figure 3 shows the results of dielectric behavior with electric-field ($E$) parallel and also perpendicular to $c$-axis (rod axis). The data is taken while warming with a rate of 0.5K/min. The dielectric constant ($\varepsilon'$) undergoes an abrupt change with $T$ in the range 60-100 K for all ν (figure 3a). This upturn occurs at about 60 K for 1 kHz, and this temperature is nearly the same as the $T_p$ observed in $\chi'$ for this frequency, as though ac χ and dielectric properties track each other. While this indicates that the origin of this dielectric feature could lie in the magnetic ordering, it is possible that Maxwell-Wagner (MW) effect [27] also contributes to such a feature in this $T$-region. The 'upturn temperature' increases strongly with ν, reaching a value of about 100 K for 100 kHz, mimicking relaxor ferroelectrics as well as MW effect [27]. However, if the feature is exclusively due to MW effect, then the magnitude of $\varepsilon'$ following the upturn should dramatically decrease as shown in Ref. 27, in contrast to present observation. This characteristic temperature also corresponds to $T_p$ in the loss part (tanδ) (figure 3b) and is found to obey the Arrhenius equation (shown inset of figure 3a). $E_a$ turns out to be about 820 K, in fair agreement with that from ac χ data, supporting the role of magnetism. Moreover, the electrical resistivity by a four-probe method could be measurable above 200 K only, due to highly insulating behavior at lower temperatures (see Supplementary Material [24]) and the value of $E_a$ derived from this data is significantly higher (> 3000 K) for two orientations (namely, the excitation current being parallel to $c$ and perpendicular to $c$) as though MW contributions bear less significant effect on the above-mentioned dielectric features. Figures 3c and 3d show the dielectric behavior for the orientation of $E \perp c$. The ν-dependence is similar to that for $E//c$ (except a small shift of the peaks towards higher temperatures), obeying the Arrhenius law (inset of figure 3c) with essentially same value for $E_a$.

The fact that this *T*-dependent dielectric glassy-like behavior appears isotropic in nature in contrast to (anisotropic) ac χ is intriguing. However, from the view of absolute values of ε' (being different for the two orientations, see figure 3), the dielectric behavior appears to be anisotropic. There could be two explanations for this: (i) It is possible that, for *E//c,* intrinsic magnetic contribution may be competing, whereas, for *E⊥c,* MW effect dominates, as a result of anisotropic magnetism and/or (ii) the dynamics of spin and electric dipoles are different. All these dielectric features are found to be robust to the application of magnetic fields [24]. For *E//c*, there is an extra shoulder around 90-120K (figure 3b) attributable to additional relaxation, which is not apparent for the other direction; this could also support anisotropic dielectric behavior.

Interference from MW contribution well below $T_N$ can be discarded due to highly insulating character (also see small values of tanδ in figures 3 and 4 below 40 K, endorsing this). Therefore, concrete evidence for the existence of magnetodielectric coupling and phenomena associated with this is obtained from isothermal data at such low temperatures. We show the changes observed in ε' (expressed in terms of $\Delta\varepsilon' = [(\varepsilon'(H) - \varepsilon'(0))/\varepsilon'(0)]$) with the variation of *H* at various temperatures in figure 4 for different geometries for 50 kHz. For the geometry *E//c//H* (Fig. 4a-d), the magnitude of overall change at higher temperatures (e.g., 30 and 40K, shown in Fig. 4c-d) is noticeably higher (as large as 8%) than that reported for isostructural $Ca_3Co_2O_6$ by Basu et al [4]. Looking at the curves for 30K and 40 K, Δε', after an initial increase, exhibits a peak and a valley (called "step") in the same *H*-range as for the step in *M(H)*, followed by an upturn and a saturation at high fields. But the hysteretic behaviour, which is weak in *M(H),* is distinctly visible in this property, more strongly at 30 K than at 40 K. It is intriguing to note that the step in these dielectric curves occurs roughly at one-third of respective high-field values, as though there is a one-to-one correlation with *M(H)* behavior. Establishing this correlation in other materials could motivate theoretical work on the role of geometrical frustration on such non-linear dielectric anomalies. There are additional qualitative changes in the shape of curves with a further lowering of temperature (see Fig. 4a-b for 2 and 10 K). Apart from the fact that the sign of Δε' changes, the magnitude also gets smaller (compare the figures 4a and 4b with the figures 4c and 4d). Hence it is difficult to attach any significance to subtle features. In any case, it is clear that there are qualitative and quantitative changes in MDE across $T_2$. Clearly, MDE tracks various magnetic ordering regimes arising from geometrical frustration. At very high temperatures above $T_N$ (say, at 150 K, not shown here), the MDE is negligible. Now, turning to the behaviour for the other geometry *E//c⊥H* (figure 4g), the curves were found to be noisy without a step in the range 20-40 kOe and with a negligible change of ε' till 140 kOe. All these results viewed together prove anisotropic MDE behaviour of this compound. For *E⊥c//H* (figure 4e and 4f), the MDE features are similar to that for *E//c//H* except a relatively smaller overall change with the variation of *H*. Thus, in terms of magnitude as well, the MDE coupling is anisotropic with respect to the direction of applied *E*. This observation is consistent with the trends observed in the values in the *T*-dependent curves in figure 3. For this orientation as well, ε' captures the magnetic field-induced transition with a hysteretic effect above $T_2$ (as in the case for *E//c*); at lower temperatures, for instance for 2 and 10K, the change in ε' is negligibly small (not shown). For *E⊥c⊥H* (see, for instance, the data for 40 K in figure 4h), MDE coupling is negligible. The tanδ also captures the features in all cases (see insets of figure 4d and 4f). In short, viewing all these data together, we can confidently state that this compound exhibits anisotropic MDE behaviour with respect to the direction of E and *H*.

Summarizing, the results, apart from clearly establishing exotic anisotropic spin-glass-like behavior of $Ca_3Co_{1.4}Rh_{0.6}O_6$, reveals the existence of anisotropic magnetodielectric coupling attributable to geometrically frustrated magnetism. The fascinating finding of a step in isothermal Δε' at about 1/3 of saturation value at high magnetic fields mimicking the one in *M(H)* is of theoretical interest and further exploration of this on other materials would advance our knowledge on the applications of antiferromagnetically coupled triangular spin systems.

Figure 1: (a, b) Dc magnetization divided by magnetic field as a function of temperature for two orientations. In (c) and (d) isothermal magnetization behavior is shown for different temperatures for *H//c* and in (e) the same is shown for *H⊥c*. Inset in (c) shows the curve for 30 K in an expanded form to highlight hysteretic effect.

Figure 2: The real ($\chi'$) and imaginary ($\chi''$) parts of ac susceptibility as a function of temperature at various frequencies and different dc magnetic fields for the orientation *H//c*. The lines are drawn through the experimental data points to serve as guides to the eyes. Inset shows Arrhenius plot. The arrows show the direction in which the curves move with increasing frequency. Some curves at high fields are very noisy and hence are not shown.

Figure 3: Dielectric constant and tanδ for $Ca_3Co_{1.4}Rh_{0.6}O_6$. The arrows show how the curves move with increasing frequency from 1 to 100kHz (1, 5, 10, 20, 30, 50, 70, and 100 kHz). Inset shows Arrhenius plot. Panels (a) and (b) are for *E//c* and figures (c) and (d) are for *E⊥c*.

Figure 4: The change in the dielectric constant ($\Delta\varepsilon' = [\{\varepsilon'(H)-\varepsilon'(0)\}/\varepsilon'(0)]$) with the variation of magnetic field for $Ca_3Co_{1.4}Rh_{0.6}O_6$. (a), (b), (c), and (d) are for *E//c//H* at 2, 10, 30 and 40 K respectively. (e) and (f) are for *E⊥c//H* at 30 and 40 K. Insets in (d) and (f) show tanδ at 40K. The data for *E//c⊥H* and *E⊥c⊥H* at 40 K are also shown in (g) and (h). The arrows indicate the direction in which *H* is varied.

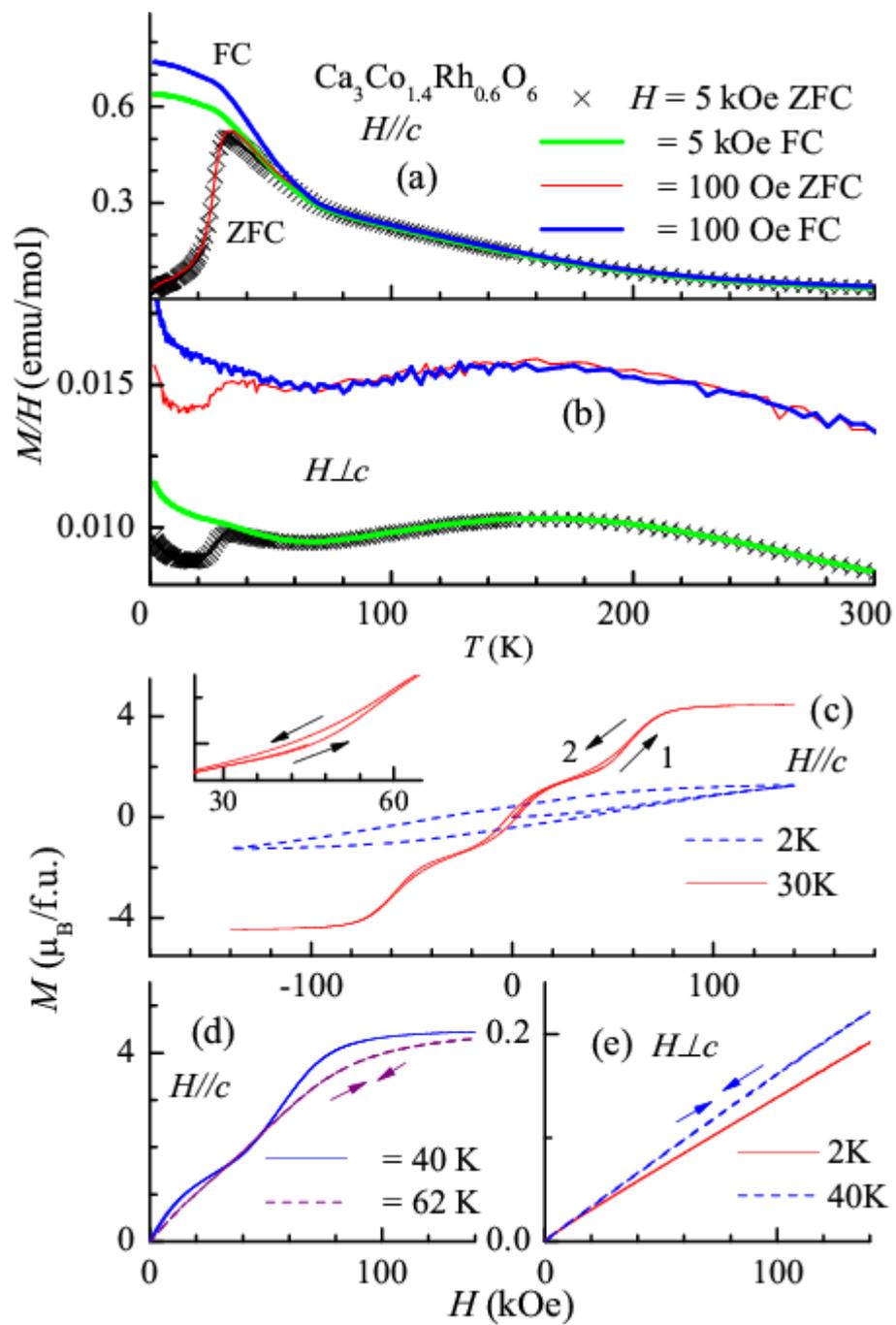

Figure 1

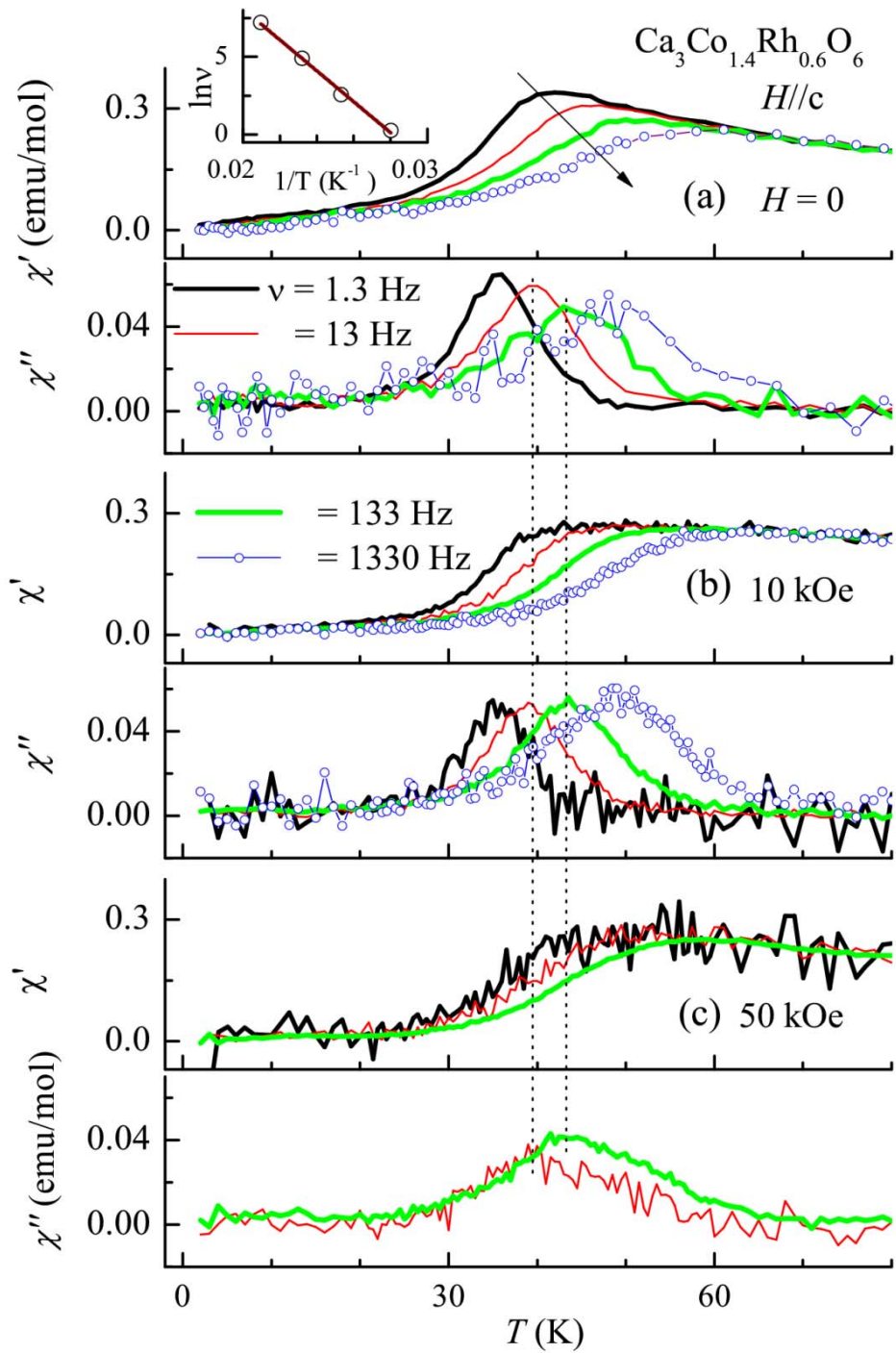

Figure 2

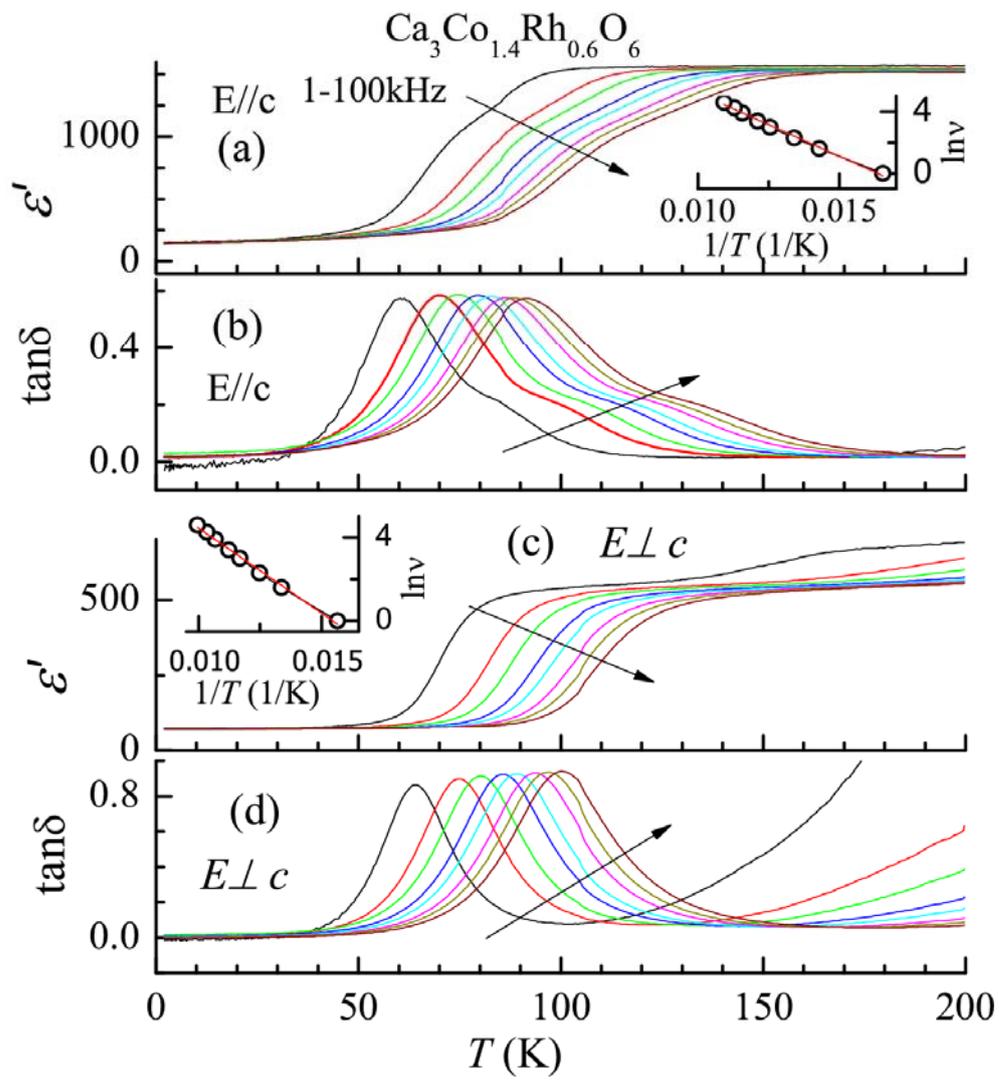

Figure 3

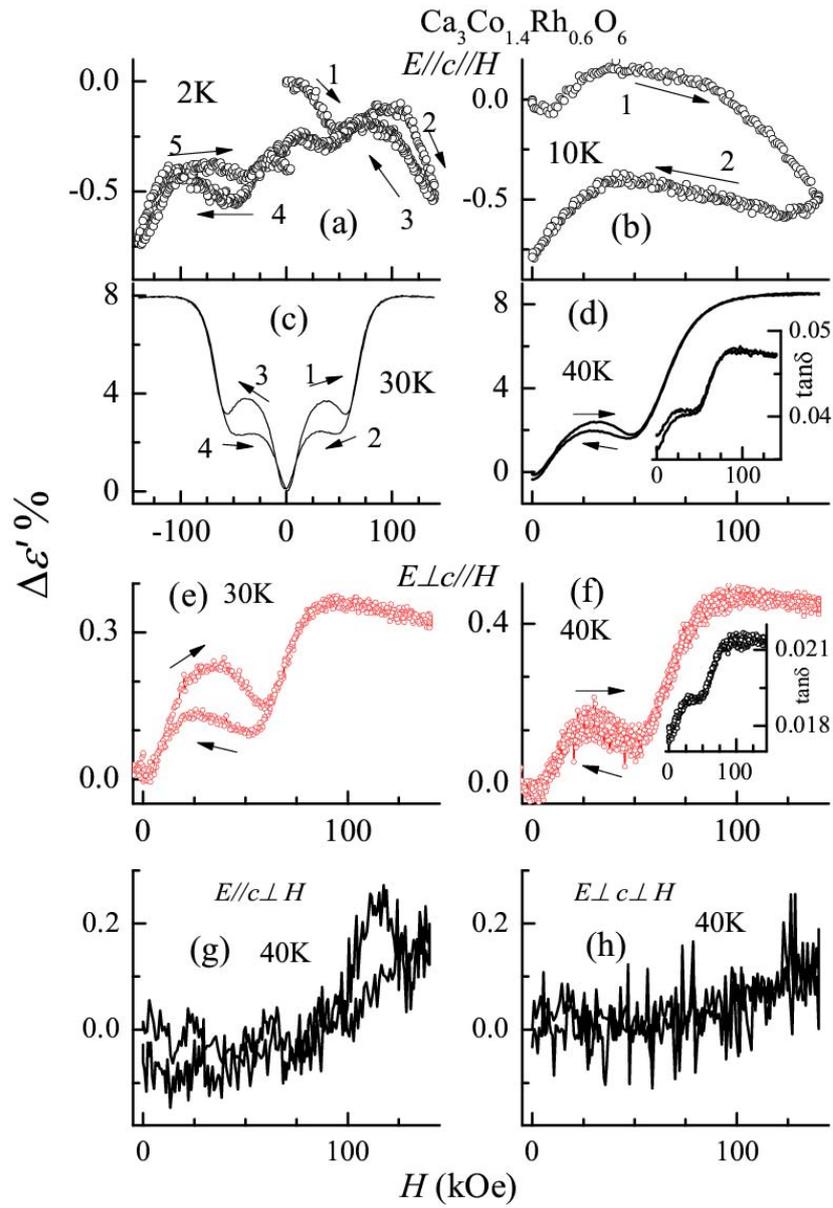

Figure 4

**Supplementary Material**

**Anisotropic magnetodielectric coupling behavior of Ca$_3$Co$_{1.4}$Rh$_{0.6}$O$_6$ due to geometrically frustrated magnetism**

Tathamay Basu, Kartik K Iyer, Kiran Singh, K. Mukherjee, P.L. Paulose and E.V. Sampathkumaran

*Tata Institute of Fundamental Research, Homi Bhabha Road, Colaba, Mumbai 400005, India*

In this Supplemental Material, we present data on the shape of single crystal (Figure S1) and sample characterization, isothermal remnant magnetization behavior (Figure S2), and dielectric and dc resistivity data in the presence of external magnetic fields (Figure S3).

About 2mm long needle-shaped hexagonal single crystals [24] were grown from a molten K$_2$CO$_3$ as prescribed by Davis et al [7] with the starting composition Ca$_3$CoRhO$_6$. The samples thus prepared were characterized by powder x-ray diffraction (XRD) on crushed single crystals. However, the XRD pattern for the single crystalline specimen is shifted to a marginally higher angle side compared to that for polycrystalline specimen (synthesized through a conventional solid state reaction route [10]), establishing that the lattice constants for the former are marginally lower ($a$= 9.184(1)Å, $c$= 10.684(1)Å for single crystals; a= 9.199(1)Å, c= 10.734(1)Å for polycrystals). In order to verify the composition, the specimens were further characterized by scanning electron microscopy. While the specimens were found to be homogeneous, the composition as determined by Energy Dispersive X-ray analysis corresponds to the formula Ca$_3$Co$_{1.4}$Rh$_{0.6}$O$_6$ very close to that reported by Davis et al [7]. This observation, viewed together with reduced lattice constants with respect to polycrystals, implies partial occupation of octahedral (Rh) site by Co ions (which are smaller in ionic radii compared to Rh).

In order to render support to the anisotropic nature of glassy-like magnetic behavior, we have also investigated isothermal remnant magnetization, $M_{IRM}$, behavior at various temperatures (5, 15 and 40 K) below and above $T_2$. For this purpose, we have cooled the specimen in the absence of any external magnetic field from 300 K to desired temperatures, switched on a field of 5 kOe and, after 5 minutes, the field was switched off. Subsequently, $M_{IRM}$ was measured as a function of time ($t$). The curves obtained are shown in figure S2. The decay of $M_{IRM}$ is rather slow for all temperatures for $H//c$. However, for $H\perp c$, the decay is faster; that is, following an initial drop, the value stays almost constant with very small magnitude (as a result of which the plots are noisy). For this orientation, the curve obtained for 40 K is similar to those for 5 and 15 K and since the data points are quite noisy at $t$= 0, we have omitted the curve for this temperature. The findings strongly endorses anisotropic glassy-like magnetism. We have also fitted the data to stretched exponential form ($M_{IRM}$(t)/$M_{IRM}$(0))= a + b exp[(t/$\tau$)$^{0.5}$]) (where a and b here are coefficients) and the values of relaxation time ($\tau$) obtained from fitting are given in the figure.

We now discuss how the glassy-like dielectric features observed as a function of temperature respond to the application of magnetic fields. A fascinating finding we have made is that the glassy features observed in $\varepsilon'(T)$ and in tan$\delta(T)$ remain the same up to 100 kOe irrespective of the orientation or the geometry. To demonstrate this, we show the data in figure S3 for $E//c//H$ for an application of 100 kOe. We observe a small shift (2K) in peak temperature for tan$\delta$ towards lower temperatures with increasing $H$, as shown in figure S3b for a particular frequency of 30 kHz. A similar change in the peak temperature is observed for $E\perp c//H$ (figure S3c). We have obtained the curves as a function of temperature when $H$ is perpendicular to the chain as well and we could not resolve any change with the variation of $H$.

In the inset of figure S3c, we also show dc electrical resistivity behavior in zero magnetic field as well as in 50 kOe for $E\perp c$. This data was obtained by two probe method using an electrometer (Keithley, 6517A), since the values are beyond measurable limit using a conventional four-probe method below about 200 K. The electrical resistance sharply increases attaining tera-ohm range with a lowering of temperature below 70 K for $E\perp c$ and so the resistance could not be measurable at further lower temperatures even by this method. Clearly, at much lower temperatures, the absolute values of resistivity are several giga-Ωcm. The behavior for $E//c$ is similar. We could not detect any variation with the application of magnetic fields even above 70 K. Clearly, in the temperature region of interest to conclude on MDE, for instance, based on figure 4 (<50 K), the material is highly insulating that intrinsic contributions to dielectric anomalies should be significant. This conclusion is supported by negligibly small values of tanδ, for instance at 40 K (see the insets in figure 4d and 4f).

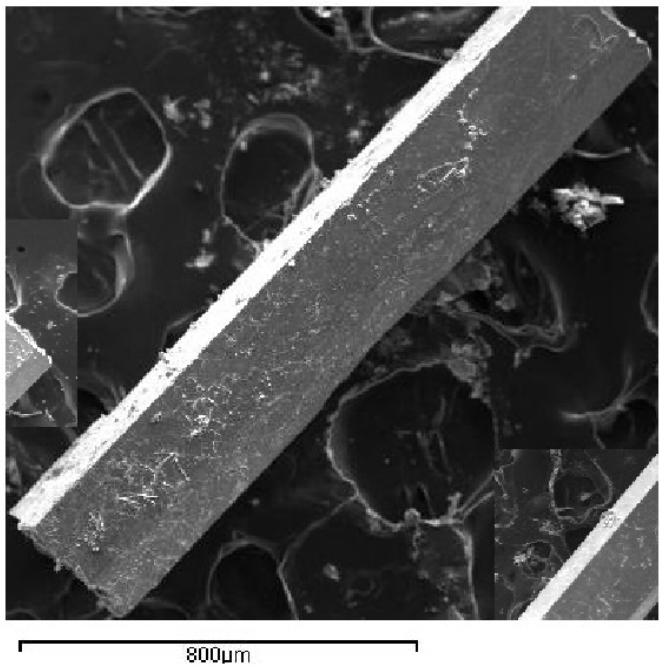

Figure S1: Typical scanning electron micrograph of the sample to reveal the crystal shape and size.

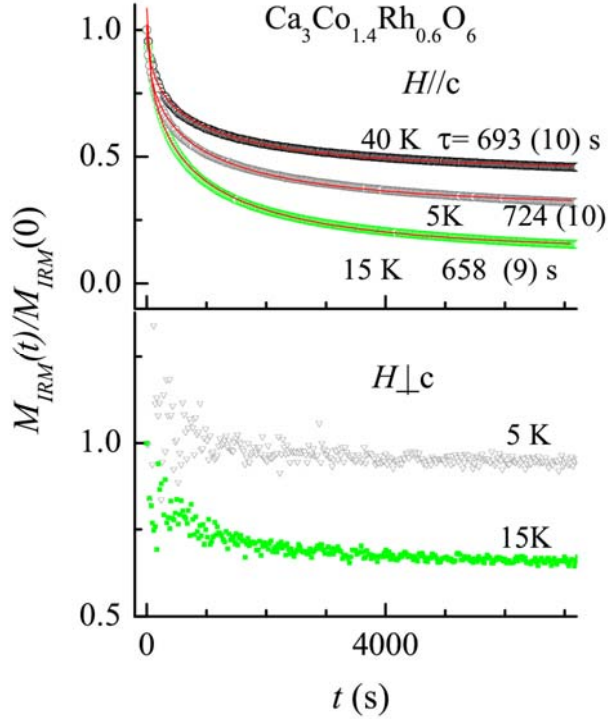

Figure S2:
Isothermal remnant magnetization behavior at 5, 15 and 40 K. The continuous lines for *H//c* are fit to the form, $M_{IRM}(t)/M_{IRM}(0)) = a + b \exp[(t/\tau)^{0.5}]$ and the relaxation times ($\tau$) obtained by fitting are also mentioned. For *H//c*, the values of $M_{IRM}(0)$ are: ~12.42, 20.73 and 12.58 emu/mol for 5, 15 and 40 K respectively; for *H⊥c*, the corresponding values are extremely small ~0.38, 0.53 and 0.33 emu/mol and so the plots are very noisy particularly neat *t*= 0 . The coefficients a and b (for *H//c*) are: ~0.3 and 0.71 for 5K, 0.12 and 0.96 for 15 K, and 0.44 and 0.58 for 40 K respectively. The error bars on these values are negligible.

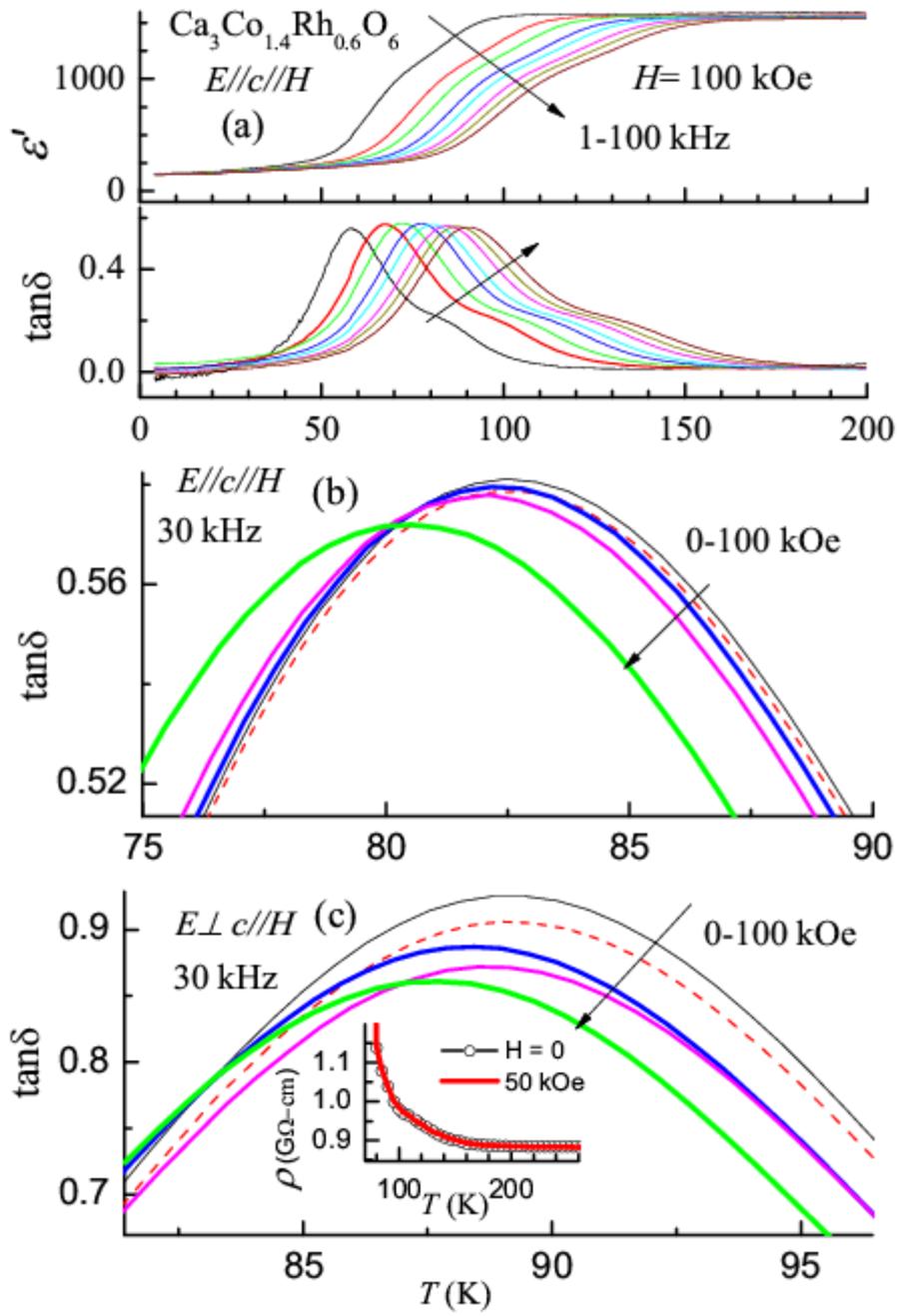

Figure S3: *(a)* Dielectric constant and tanδ for *E//c//H* for 100 kOe; *(b)* and *(c)* show loss factor for different fields (0, 10, 30, 50 and 100 kOe) measured with 30 kHz for two different geometries *E//c//H* and *E⊥c//H* respectively. Inset in (c) show the dc electrical resistivity as a function of temperature for *H*=0 and 50 kOe.